\documentclass[fleqn,usenatbib]{aa}
\usepackage[breaklinks=true]{hyperref}
\usepackage[utf8]{inputenc}
\usepackage[varg]{txfonts}
\usepackage{color}
\usepackage{xcolor}
\usepackage{lipsum}
\usepackage[english]{babel}
\usepackage{fontenc,amssymb}
\usepackage{graphicx}
\usepackage{amsmath}
\usepackage[normalem]{ulem}

\graphicspath{{Figures/}}

\newcommand{\SiII}{\hbox{{\rm Si}{\sc \,ii}}}
\newcommand{\SII}{\hbox{{\rm S}{\sc \,ii}}}

 \newcommand{\mpy}{\hbox{M$_{\odot}$~yr$^{-1}$}}

\newcommand{\spectranumber}{8690}
\newcommand{\um}{\textmu m}
\newcommand{\kms}{\hbox{km~s$^{-1}$}}

\begin{document}
\title{PDRs4All VII. The 3.3 \um\ aromatic infrared band as a tracer of physical properties of the ISM in galaxies~\thanks{Based on JWST ERS program \#1288}}

\author{
    Ilane~Schroetter \inst{1} \and
        Olivier~Bern\'e \inst{1} \and
        Christine~Joblin \inst{1} \and
        Amélie~Canin \inst{1} \and
        Ryan~Chown \inst{2,3} \and
        Ameek~Sidhu \inst{2,3} \and
        Emilie~Habart \inst{8} \and
        Els~Peeters \inst{2,3,4} \and
        Thomas~S.-Y.~Lai \inst{9} \and
        Alessandra~Candian \inst{5} \and
        Shubhadip~Chakraborty \inst{6,7} \and
        Annemieke~Petrignani \inst{10} 
        }
\institute{Institut de Recherche en Astrophysique et Plan\'etologie, Universit\'e Toulouse III - Paul Sabatier, CNRS, CNES, 9 Av. du colonel Roche, 31028 Toulouse Cedex 04, France \label{inst1} \and
{Department of Physics \& Astronomy, The University of Western Ontario, London ON N6A 3K7, Canada}\and
{Institute for Earth and Space Exploration, The University of Western Ontario, London ON N6A 3K7, Canada}\and
{Carl Sagan Center, SETI Institute, 339 Bernardo Avenue, Suite 200, Mountain View, CA 94043, USA}\and
Anton Pannekoek Institute for Astronomy, University of Amsterdam, The Netherlands \and
Department of Chemistry, GITAM School of Science, GITAM Deemed to be University, Bangalore, India \and
Institut de Physique de Rennes, UMR CNRS 6251, Universit{\'e} de Rennes 1, Campus de Beaulieu, 35042 Rennes Cedex, France \and
Institut d'Astrophysique Spatiale, Universit\'e Paris-Saclay, CNRS,  B$\hat{a}$timent 121, 91405 Orsay Cedex, France \and
IPAC, California Institute of Technology, Pasadena, CA, USA \and
Van 't Hoff Institute for Molecular Sciences, University of Amsterdam, The Netherlands
}

\label{firstpage}
\abstract
{Aromatic infrared bands (AIBs) are a set of broad emission bands at 3.3, 6.2, 7.7, 8.6, 11.2, and 12.7 $\mu$m, seen in the infrared spectra of most galaxies. With JWST, the 3.3 $\mu$m AIB can in principle be detected up to a redshift of $\sim 7$. Relating the evolution of the 3.3 $\mu$m AIB to local physical properties of the ISM is thus of paramount importance. 

{By} applying a dedicated machine learning algorithm to JWST NIRSpec observations of the Orion Bar photodissociation region obtained as part of the PDRs4All Early Release Science (ERS) program, we extracted two template spectra 
{capturing the evolution of the AIB-related emission in the 3.2-3.6~\um\ range, which includes the AIB at 3.3~\um\ and its main satellite band at 3.4~\um.}

{In the Orion Bar,} we analyze the spatial distribution of the templates and their relationship {with the fluorescent emission of H$_2$ in the near infrared.}

We find that one template (``AIB$_{\rm Irrad}$'')  traces regions of neutral atomic gas with strong far-UV fields, while the other template (``AIB$_{\rm Shielded}$'') corresponds to shielded regions with lower FUV fields and a higher molecular gas fraction.
We then show that these two templates can be used to fit the NIRSpec AIB-related spectra of
nearby galaxies. The relative weight of the two templates (AIB$_{\rm Irrad/Shielded}$) is a tracer of the radiative feedback from massive stars on the ISM. We derive an estimate of 
AIB$_{\rm Irrad/Shielded}$ in a $z=4.22$ lensed galaxy, and find that it has a lower value than for local
galaxies. This pilot study illustrates how a detailed analysis of AIB emission in nearby regions
can be used to probe the physical conditions of the extragalactic ISM. }


\titlerunning{The 3.3 \um\ AIB as a tracer of physical properties of the ISM in galaxies}                
\maketitle
\section{Introduction}
Photodissociation regions (PDRs) are region of the interstellar medium where the 
far-UV photons (6\,$<$ E $<$ 13.6~eV) from massive stars strongly influence the
dust and gas. 
This deposition of energy results in the dissociation of molecules and heating 
of the gas and dust. PDRs cool through emission in the infrared (IR). In the mid-IR wavelength range ($3-28$~\um), classic spectral signatures of PDRs are 
i) continuum emission attributed to dust grains, ii) H$_2$\ emission lines, 
iii) emission lines from neutral atoms and ions ([\SII], [\SiII], etc.), and
iv) aromatic infrared bands (AIBs), 
 which are broad emission features generally attributed to fluorescent emission of large carbonaceous molecules, i.e. polycyclic aromatic hydrocarbons (PAHs). 
The most prominent AIBs are found at wavelengths of 3.3, 6.2, 7.7, 8.6, 11.2, and 12.7 \um.
Perhaps the main observational fact is that PAHs are ubiquitously observed in the interstellar medium (ISM) 
of star-forming galaxies \citep[SFGs, e.g.][]{DraineB_07, LiA_20, SandstromK_23}, including 
at high redshift \citep[e.g.][]{RiechersD_14, McKinneyJ_20}. 
PAHs are believed to play a major role in the physics and chemistry of PDRs, and, notably, in heating the gas
via the photoelectric effect \citep[e.g.][]{BakesE_94,WeingartnerJ_01, BerneO_22}.

{Here we focus on the emission in the 3.2-3.6~\um\ range, which includes the AIB at  3.3~\um\ and less-prominent neighboring bands, in particular one at 3.4~\um. While the emission at  3.3~\um\ is attributed to aromatic C-H stretching vibrations, the emission at  3.4~\um\ is attributed to aliphatic C-H stretching vibrations \citep[e.g.][]{AllamandolaL_89, joblin96, YangXJ_16}.}
These bands are also seen in galaxy spectra \citep[i.e.][ and references therein]{LiA_20}.

\citet{KimJ_12} as well as \citet{LaiT_20} showed that the 3.3 \um\ emission can be a reasonable star formation (SF) indicator and 
\citet{RigopoulouD_21} demonstrated that PAH intensity ratios could be used to probe physical conditions of the ISM of galaxies and thus differentiate between normal SFGs and galaxies hosting an AGN. Using near-IR spectra of M82 observed by AKARI, \citet{YamagishiM_12} found that {the spatial evolution of the spectra can be explained by two main components: the AIB at 3.3 \um\ and the aliphatic satellite band at 3.4 \um.} 
{The ratio between the intensity of the 3.3 and 3.4~\um\ bands seems to increase the closer to the galaxy AGN the observations are.} 

{Observations with the James Webb Telescope (JWST) opens up new possibilities for using these signatures to trace the physical conditions in galaxies.}
Indeed, first  observations with the JWST have shown that {the AIB at 3.3 \um\ and its satellite at 3.4 \um} are observed and bright in galaxies of the 
nearby universe \citep[e.g.][]{GOALS_JWST_vv114, GOALS_JWST_iizw96, GOALS_JWST_ngc7469, LaiT_23} and up to a redshift of 4 \citep[the lensed galaxy SPT0418-47][]{SpilkerJ_23}, where the only detected spectral feature in the galaxy spectrum is {the AIB at 3.3 \um} observed with JWST/MIRI MRS. 
In principle, this AIB could be detected up to a redshift $z\approx7$ with MIRI MRS. {It is therefore useful to understand how this emission is linked to local physical properties.}

In this article, using JWST-ERS NIRSpec observations of the Orion Bar (program ID \# 1288, \citet{PDRs4All_22}), we extract elementary spectra using a machine learning approach based on non-negative matrix factorization (NMF), following previous studies \citet{Berne_O_07, Foscino_19}. 
The article is composed as follows: 
we describe the data in \S~\ref{data} and in \S~\ref{section:formalism}, we detail the process of extracting the AIB templates {in the near-IR (3.2-3.7~\um) range} for the Orion Bar. 
We then compare these templates with the ones from \citet{Foscino_19}.
In \S~\ref{sec:spatial_dist_templates}, we analyse the spatial distribution of the templates and establish the relationship between these templates and the ISM properties in \S\ref{section:ism_properties}.
We then use the templates to analyze galaxy spectra in \S~\ref{section:galaxies} and we extend {the method to the spectra of spatially resolved galaxies in \S~\ref{subsection:resolved_fitting}.}.
Finally, conclusions are presented in \S~\ref{section:conclusion}.
Throughout the paper, we use a 737 cosmology ($H_0=70$~\kms, $\Omega_{\rm m}=0.3$, and $\Omega_{\Lambda} = 0.7$) and all errors are given at 1$\sigma$ unless stated otherwise.

\section{Data}
\label{data}

\subsection{Orion Bar}

The observations were performed with the JWST-NIRSpec Integral Field Unit (IFU) as part 
of the PDRs4All ERS program \citep{PDRs4All_22} on September 10 2022. 
They were reduced using the JWST pipeline version 1.9.4 with Calibration 
Reference Data System (CRDS) context file jwst\_1014.pmap. 
For this study, we only use the spectral cube corresponding to the F290LP filter which 
spans from $\sim2.9$ to 5~$\mu$m. This data cube is a combination of 9 pointings, forming 
a mosaic spanning across the Orion Bar. The detailed data reduction together with a general 
analysis of the line emission is described in \citet{PeetersE_23} and in \citet{ChownR_embarrassment23} {where} a first description of the PAH emission is also discussed.  

\subsection{Galaxies}

At the time of writing this paper, four galaxies from the ERS program 
\#1328\footnote{Details on this program can be seen on their website 
\href{https://goals.ipac.caltech.edu/}{https://goals.ipac.caltech.edu/}} 
have been observed with NIRSpec. Those galaxies are, namely, VV114 
\citep[][]{GOALS_JWST_vv114, GOALS_JWST_vv114_b, GOALS_JWST_vv114_c}, NGC7469 
\citep[][]{GOALS_JWST_ngc7469, GOALS_JWST_ngc7469_b, GOALS_JWST_ngc7469_c, GOALS_JWST_ngc7469_e, LaiT_23, BianchinM_23}, 
IIZw96 \citep[][]{GOALS_JWST_iizw96} and NGC3256 nucleus 1 and nucleus 2.
We retrieved from MAST the level 3 NIRSpec IFU cubes for these five pointings 
covering filters 100, 170 and 290LP. 
For each pointing, we extracted from the cubes the average spectrum 
using elliptical apertures (see Table~\ref{table:galaxies_extraction}). 
As most of the galaxies observed have saturated spectra near their center, we exclude the saturated area when extracting each galaxy spectrum.

We also include data for the M82 galaxy from the GO program id \# 2677. 
These are NIRSpec MSA observations of the nucleus and disk edge with the F290LP filter. Here we use the data for the nucleus only since it provides the best detection of the {AIB at 3.3~\um\ and its satellite band at 3.4~\um.} 
We obtained a mean spectrum by averaging the spectra obtained in all 
shutters. 

In addition, another ERS program (id \# 1355) released their MIRI MRS reduced data of a redshift $z\approx4$ galaxy, {called SPT0418-47, in which the 
AIB at 3.3~\um\ is present \citep{SpilkerJ_23}.}
For this specific galaxy, we follow the reduction process {provided by the authors} 
in order to extract a unique
average spectrum of this lensed source for all channels of MIRI.

In total, we thus obtain 7 average spectra covering the {range of the emission bands at 3.3 and 3.4 \um} 
in galaxies near and far. 

\begin{table*}
\centering
\caption{Aperture parameters that were used to extract NIRSpec spectra from nearby galaxies}
\label{table:galaxies_extraction}
\begin{tabular}{lccccc}
\hline
Galaxy name    & Center RA & Center DEC & radius ($l\times h$) & radius$_{\rm sub}$ &PA \\
(1)   &  (2)   & (3) & (4) & (5) & (6) \\
\hline
IIZw96          & 20:57:24.376 & 17:07:39.7   & 1.27$\times$0.75 & $\cdots$ & 25 \\
VV114           & 01:07:47.530 & -17:30:25.30 & 0.51$\times$0.38 & 0.13$\times$0.08 & 33 \\
NGC7469         & 23:03:15.633 & 8:52:26.21   & 0.79$\times$0.79 & $\cdots$  & 0 \\
NGC3256 nucl1   & 10:27:51.231 & -43:54:14.00 & 0.34$\times$0.34 & 0.06 & 0 \\
NGC3256 nucl2   & 10:27:51.218 & -43:54:19.42 & 0.38$\times$0.24 & 0.06 & 20 \\
\hline
\end{tabular}\\
{
(1) Galaxy name;
(2) RA of the ellipse center (J2000);
(3) DEC of the ellipse center (J2000);
(4) Radius of ellipse in kpc (semi-major and semi-minor axes);
(5) Radius of subtracted ellipse in kpc;
(6) Ellipse position angle (in degrees, same for both apertures).
}
\end{table*}

\begin{figure}[ht!]
  \centering
  \includegraphics[width=8cm]{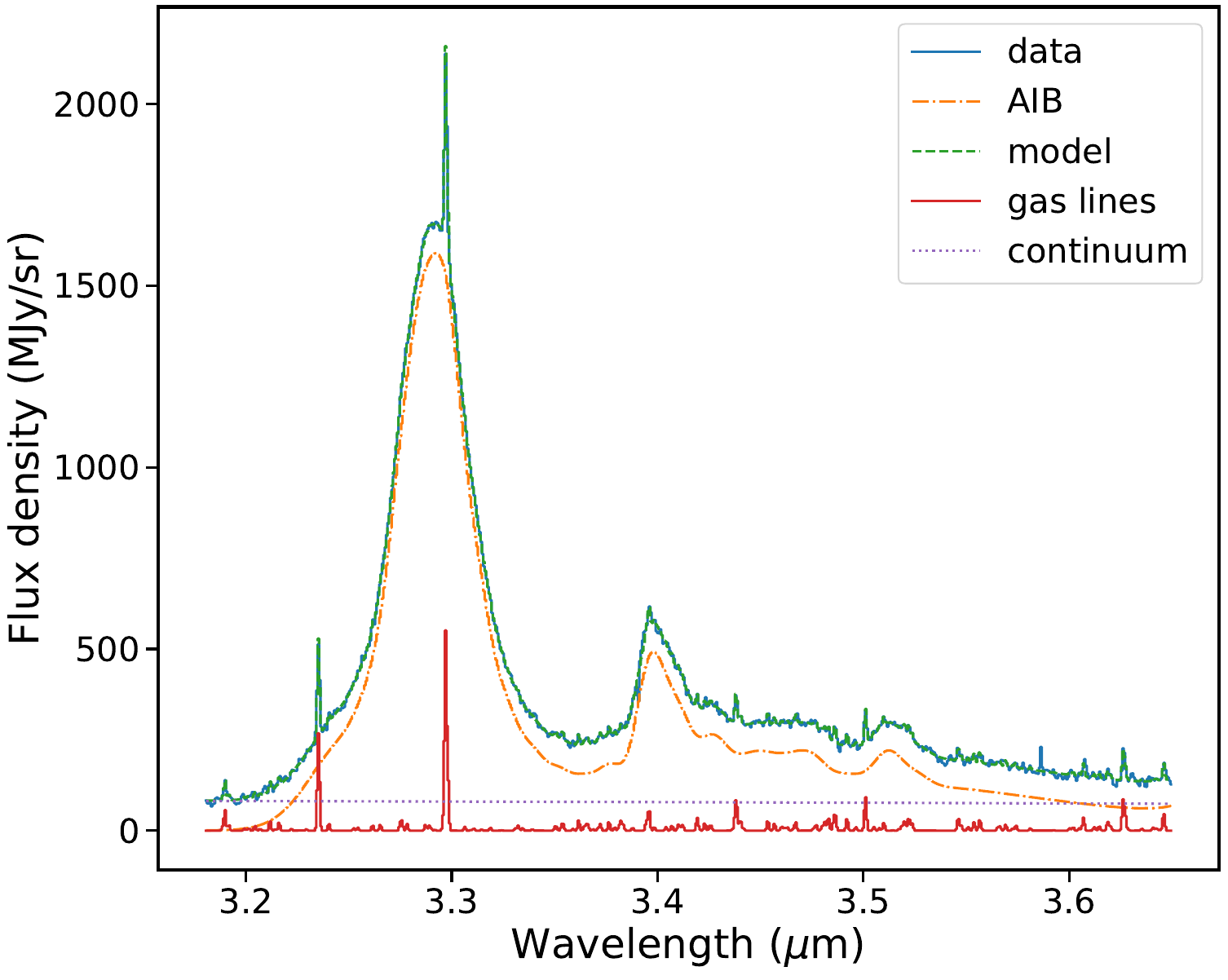}
  \caption{
  Example of a spectrum of a spaxel of the Orion Bar NIRSpec mosaic fitting. Like the other  \spectranumber \, spectra, this spectrum was extracted using an aperture with a radius of 1 pixel and was fit using a linear combination of Gaussians ({AIB-related bands} in orange), gas lines (in red) and continuum (in purple). 
  The model can be seen in dashed green on top of the data shown in blue.
  }
  \label{fig:fit}
\end{figure}

\begin{figure*}[h!]
  \centering
  \includegraphics[width=18cm]{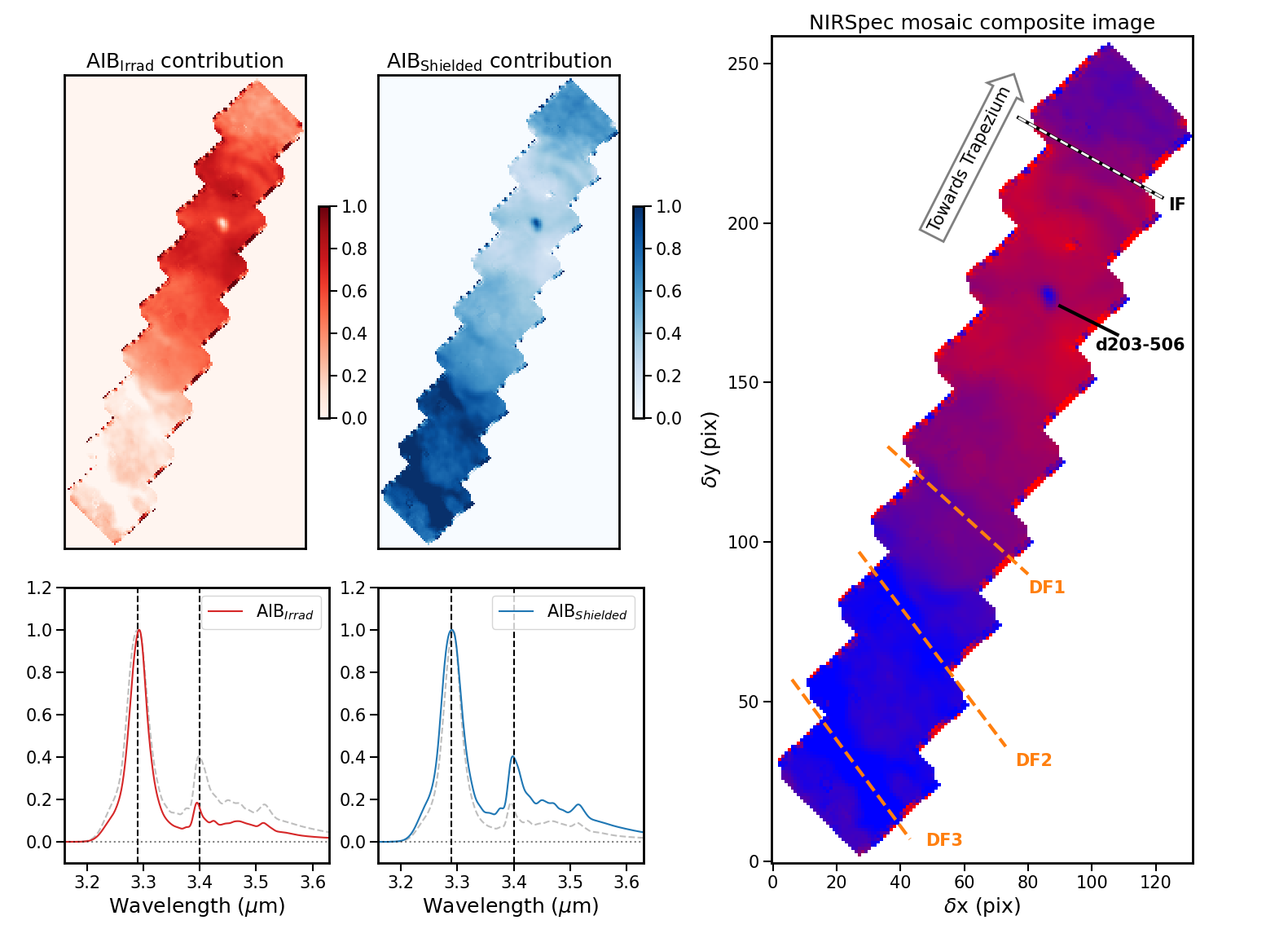}
  \caption{
  \textit{Left column: }AIB$_{\rm Irrad}$ map on top and the corresponding normalized template spectrum of the AIB$_{\rm Irrad}$ component.
  \textit{Middle column: }same as left column but for the AIB$_{\rm Shielded}$ template.
  Note that AIB$_{\rm Irrad}$ has a more prominent 3.3~\um\ band (and a slightly redder peak) than AIB$_{\rm Shielded}$ which has a stronger 3.4~\um\ band. 
  For a quick comparison between both templates, on each bottom panel, the complementary template is shown in dashed gray lines.
  \textit{Right column: }a composite image of both AIB$_{\rm Irrad}$ (red) and AIB$_{\rm Shielded}$ (blue) contributions.
  The ionization front (IF) as well as the dissociation fronts (DF1, 2 and 3) are represented by dashed white and orange lines, respectively \citep[defined in ][]{HabartE_imaging23, PeetersE_23}.
  }
  \label{fig:maps}
\end{figure*}

\section{Template NIRSpec AIB-related spectra in the Orion Bar}
\label{section:formalism}

{\subsection{Extracting template spectra from the observed data cube}}
Our aim is to extract a set of representative elementary
(``template'') spectra from the observed NIRSpec spectral cubes of the Orion Bar. To do so, we follow the decomposition method presented in \citet[][]{Boubou_21} and \citet{Foscino_19}. Details of the method are 
provided in these two references but we briefly describe them here. First, the 3D NIRSpec data cube $C(\alpha, \delta, \lambda)$ is transformed into a 2D matrix:
$X(m,\lambda)$, with $dim(C)=(m_\alpha, m_\delta, n_\lambda)$ and $dim(X)=(m, n_\lambda)$, $m=m_\alpha \times m_\delta$.
In this study, we consider a wavelength range between 3.18 $< \lambda < 3.65$~\um\ 
which covers {the AIB at 3.3~\um\ and its satellite bands, in particular the major one at 3.4~\um.} 
Then, we filter out low SNR spectra in $X$
using cutoffs in flux to remove potential spikes and saturated spectra in this wavelength range.   
Those cutoffs are the following: in each spectrum, intensities larger than $4\times 10^5$ MJy~sr$^{-1}$
are set to $10^{-16}$. This cut-off value is chosen to be larger than any real emission line in our data, and 
corresponds to spikes present in the data.
{We then select only those spectra 
whose sum of values is greater than 1} 
(this removes spectra which 
contain no information at the border of the field of view). 
The selected spectra constitute the lines of {the matrix} $X$, which has dimensions 
(\spectranumber,706) i.e 706 spectral points and \spectranumber\ spatial positions. 
Spectra in $X$ are then fitted using a combination of synthetic components ({AIB-related bands}, gas lines and continuum) in order to {remove the contribution from gas lines and dust emission and extract the emission related to the AIBs} 
(see \citealt{Foscino_19} and Fig.~\ref{fig:fit}). 
We define the matrix $X^*$ whose lines are these \spectranumber\ pure AIB fitted spectra (e.g. orange curve 
in Fig.~\ref{fig:fit}). 
The next step consists in identifying the dimension of the subspace spanned by the data in $X^*$. This is done as in \citet{Boubou_21}, by 
analysis of the eigenvectors of the covariance matrix. Using this method we find that this dimension is $r=2$. 
We then apply non-negative matrix factorization  (NMF \citealt{LeeD_nmf}) to $X^*$ with $r=2$, which provides the matrices $W$ and $H$ such that $X^* \approx WH$.
Following \citet{Boubou_21}, the NMF algorithm is initialized using the results of the MASS algorithm applied to $X^*$. 

\subsection{Extracted AIB-related spectra} 
\label{sec:pahtat_results}
The $r=2$ extracted AIB-related spectra are presented in Fig.~\ref{fig:maps}. The two spectra are similar, however they show some specific differences. The first template spectrum (AIB$_{\rm Irrad}$ hereafter) in red in Fig.~\ref{fig:maps} is dominated by a strong band at 3.294~$\mu$m, and a weak band at 3.393 $\mu$m. An underlying plateau is present between $\sim 3.36-3.53$, 
but relatively weak compared to the 3.294 band. 
The second template spectrum (AIB$_{\rm Shielded}$ hereafter)  in blue on Fig.~\ref{fig:maps} shows a prominent band at 3.290 $\mu$m, which is broader than the same band in the AIB$_{\rm Irrad}$ spectrum. The 3.4~$\mu$m emission feature in AIB$_{\rm Shielded}$ is also much stronger. 
The template AIB$_{\rm Shielded}$ also shows a more prominent plateau in the 3.36--3.53 $\mu$m range. 

For each template, we estimate the 3.4 to 3.3~$\mu$m band intensity ratios.
The 3.3~$\mu$m band intensity is obtained between 3.24 and 3.35~$\mu$m and 
the 3.4~$\mu$m band between 3.38 and 3.42~$\mu$m. 
For each band, we subtract a linear continuum between the integration ranges to get only the band emission.
The 3.4/3.3 integrated intensity band ratio is 0.04 for template AIB$_{\rm Irrad}$, and 0.10 for template AIB$_{\rm Shielded}$. 
These values correspond well to the extremes of this band ratio as derived from NIRSpec observations of NGC 7469 \citep{LaiT_23}.
As mentioned before, the 3.3 $\mu$m band is associated with aromatic C-H stretch emission, while the 3.4 $\mu$m band is attributed to aliphatic C-H emission \citep[][]{AllamandolaL_89}, {the fraction of aliphatics is found to be relatively minor (typically one methyl sidegroup per PAH for AIB$_{\rm Shielded}$ following \cite{joblin96} and \cite{Yang2016}. { If the carriers of the 3.4~$\mu$m band are surhydrogenated PAHs rather than alkylated PAHs \citep{bernstein1996hydrogenated}, then an even smaller amount of CH aliphatic bonds is required to explain the observed 3.4 to 3.3~$\mu$m band intensity ratios \citep{Yang2020}.  In all cases,} the AIB$_{\rm Irrad}$ template corresponds to the emission of PAHs {with very few attached aliphatic C-H bonds.
Because the latter are more easily photodissociated than aromatic C-H  bonds upon UV irradiation \citep{Marciniak2021}, the AIB$_{\rm Irrad}$ spectrum can be regarded as corresponding to more UV processed PAHs with respect to the AIB$_{\rm Shielded}$ spectrum. On the opposite, the species emitting the AIB$_{\rm Shielded}$ spectrum have been less exposed to UV irradiation \citep{joblin96, Pilleri_15, ChownR_embarrassment23, PeetersE_23}, which was also recently seen in NGC7469 by \cite{LaiT_23}.}
In Figure~\ref{fig:template_comparison}, we present both templates,
together with two templates from \citet{Foscino_19} in the same wavelength range, ordered by visual ratio of the 3.3 over 3.4~\um\ bands. The AIB$_{\rm Shielded}$ template is very close to the neutral PAH$^0$ template of \citet{Foscino_19} - yet with a slightly stronger 3.4 band, thus confirming that it corresponds to regions that are rather shielded from the UV or denser in which more pristine material is exposed to UV photons. The AIB$_{\rm Irrad}$ template 
is close to the PAH$^X$ template -yet with slightly more emission at 3.4~\um, which corresponds to the {most} irradiated environments.

\subsection{Spatial distribution of the extracted templates}
\label{sec:spatial_dist_templates}
To complement this interpretation, we derived the spatial distribution of templates AIB$_{\rm Irrad}$ and AIB$_{\rm Shielded}$ {spectra} in the Orion Bar. To do this, we {used} a linear combination of the two template AIB-related spectra + gas lines + continuum to {fit} each of the observed spectrum at each spatial position in the NIRSpec cube (see details of the fitting method in \citealt{Foscino_19}). The respective contribution of the two AIB template spectra, {which results from this fit}, is shown in Fig.~\ref{fig:maps}.  
The template AIB$_{\rm Irrad}$ dominates in the Northwest of the field of view, that is, in regions closer to the massive Trapezium stars, while template AIB$_{\rm Shielded}$ is found to be mostly present to the Southeast of the field of view. AIB$_{\rm Shielded}$ is also found to follow some dense structures present in the field of view of the Orion Bar, that is the d203-506 protoplanetary disk (Berné et al. 2023) and several dense dissociation fronts seen in H$_2$ with NIRCam \citep[][]{HabartE_imaging23}. 
{The observed spatial distribution for templates AIB$_{\rm Irrad}$ and AIB$_{\rm Shielded}$ spectra is compatible with the photo-chemical scenario discussed above, namely AIB$_{\rm Irrad}$ traces the regions that are more exposed to UV photons and AIB$_{\rm Shielded}$ is more characteristic of denser and UV shielded regions. 
In Figure~\ref{fig:maps}, we note however that there is an excess in AIB$_{\rm Shielded}$ beyond the ionization front in the HII region. This is most likely because the AIB emission in this upper corner of the map emanates from the background face-on PDR nebula and not the HII region itself \citep[see Fig. 5 of ][]{HabartE_imaging23}.}

\begin{figure}[ht!]
  \centering
  \includegraphics[width=8cm]{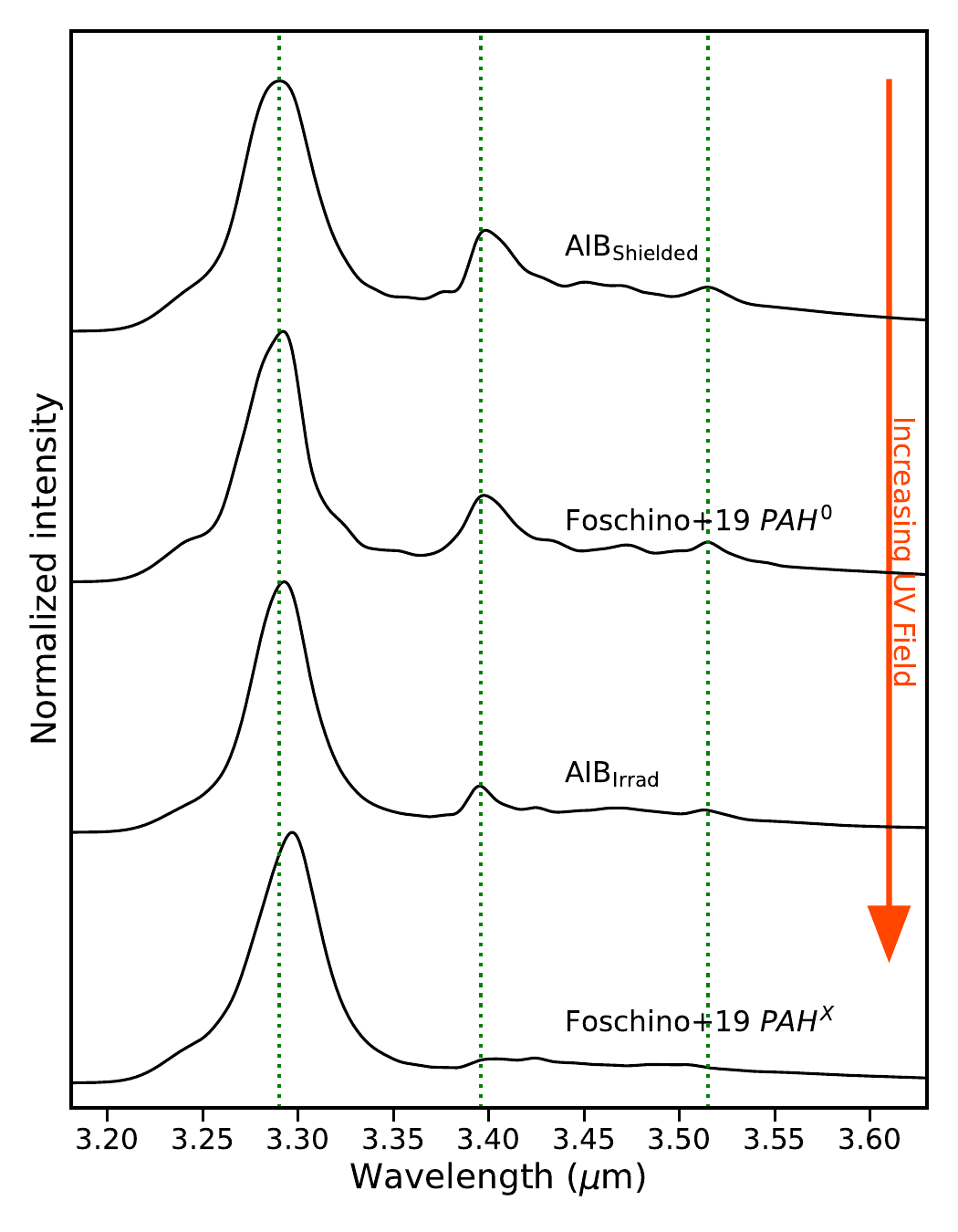}
  \caption{Comparison of AIB$_{\rm Shielded}$ (top) and AIB$_{\rm Irrad}$ (third position from top) with two templates of \citet{Foscino_19}, namely $PAH^0$ (second position from top) and $PAH^X$ (bottom). {The order from top to bottom follows the increase in UV field intensity.}
 }
  \label{fig:template_comparison}
\end{figure}

\subsection{Relationship between the template AIB-related spectra and PDR physical properties}
\label{section:ism_properties}

In order to assess the relationship between these templates and physical conditions in more detail, 
we extract several quantitative parameters on each spaxel of the Orion Bar NIRSpec field of view. 
We derive AIB$_{\rm Irrad/Shielded}$, the ratio of integrated intensity of the AIB$_{\rm Irrad}$ template over the integrated intensity of the AIB$_{\rm Shielded}$ template. 
{ We also extracted, for all pixels, the integrated intensities of 
the pure rotational H$_2$ 0--0 S(9) line 
at $\lambda = 4.695$~\um\,, the H$_2$ 1--0 S(1) ro-vibrational line at $\lambda= 2.122$~\um\,, 
the Pfund Hydrogen recombination line at $\lambda=3.297$~\um\, and the O~\textsc{i} line at
$\lambda=1.317$~\um\,. This latter line emission results from UV induced fluorescence,  
and can be used to derive the intensity of the UV radiation field $G_0$  
(see \citet{PeetersE_23} and references therein). We thus follow the approach from 
these authors to estimate the distribution of $G_0$ (in units of the \citet{Habing68} field) 
within the NIRSpec filed of view. }
Finally, we use the extracted integrated intensity of the {AIB-related} emission 
between 3.2 and 3.7 \um\ from \citet{PeetersE_23}, written I$_{\rm AIB}$. 

{ In Figure~\ref{fig:ratio}, we plot  AIB$_{\rm Irrad/Shielded}$ as a function of the 
ratios between AIB emission and the H$_2$ line intensities, 
the ratio of the Pfund-$\delta$ Hydrogen recombination line at $\lambda=3.297$~\um\, to the 
 H$_2$ pure rotational line at $\lambda = 4.695$~\um\,, and the intensity of the UV radiation 
 field derived from the O~\textsc{i} line at $\lambda=1.317$~\um\,.
Strong variations of the AIB$_{\rm Irrad/Shielded}$ ratio are observed, 
from  $\sim 10^{-3}$ to $\sim 5$. Some outliers are present in this diagram with values $>10$, corresponding 
to the edge of the NIRSpec mosaic where the fit is poor. 
All ratios on the X-axis of the four panels in Figure~\ref{fig:ratio} have been chosen 
to increase in more irradiated regions. 
AIB$_{\rm Irrad/Shielded}$ thus appears to be large in regions where the UV field is the strongest and
where H$_2$ is photodissociated (but not the AIB carriers). AIB$_{\rm Irrad/Shielded}$  is smaller in regions 
of lower UV field, where H$_2$ can form and thus the {AIB to H$_2$} ratio is {smaller}.
Similarly, AIB$_{\rm Irrad/Shielded}$ is larger 
in regions where the Pfund-$\delta$ (inonized gas) over H$_2$ (warm molecular gas) ratio is large. 
Finally, this trend is also observed with radiation field, that is, AIB$_{\rm Irrad/Shielded}$ increases 
with the intensity of the radiation field as derived from the O~\textsc{i} line. 
More specifically, the points with the top 10$\%$ AIB$_{\rm Irrad/Shielded}$ ratios have a median 
G$_0=2.0\times10^{4}$, while those with the bottom 10$\%$ AIB$_{\rm Irrad/Shielded}$ratio have a 
median G$_0=4.9\times10^{3}$. 
We note that this approach only allows to probe the intensity of the radiation field, 
while the hardness could also play a role on the AIB derived ratio (see \citealt{yang2023aliphatics}). 
The effect of hardness could be tested by observing several PDRs illuminated by 
stars with various masses.}
The increase of AIB$_{\rm Irrad/Shielded}$ is {steep} in the range where 
the AIB to H$_2$ ratio is comprised between a few $10$s and $2 \times 10^{2}$, and then much 
shallower beyond this latter value creating a "shoulder" in the diagrams upper panels 
of Fig.~\ref{fig:ratio}, this shoulder corresponds to spectra at the dissociation fronts present 
in the Orion Bar (DF1, DF2, DF3, see Fig.~\ref{fig:maps}).
Overall, these results support the interpretation in Sect.~\ref{sec:pahtat_results} and 
Sect.~\ref{sec:spatial_dist_templates} related to the processing of the AIB carriers 
with UV field inside the PDR. 

\begin{figure*}[ht!]
  \centering
  \includegraphics[width=18cm]{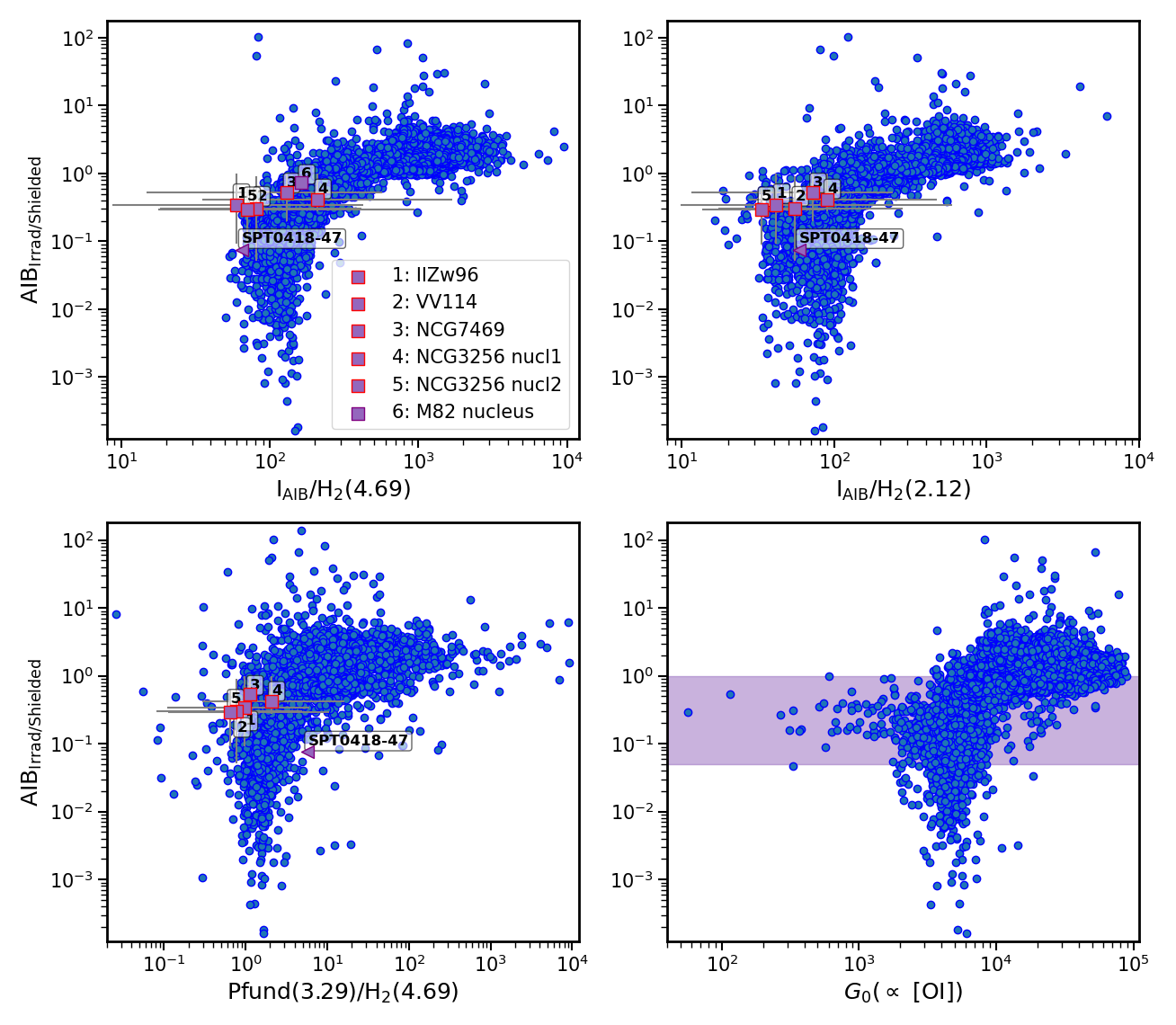}
  \caption{
  AIB$_{\rm Irrad/Shielded}$ as a function of the I$_{AIB}/$H$_2$  flux ratio for the  rotational H$_2$ 0--0 S(9) emission line at $\lambda\approx4.69~\mu$m ({top-}left panel) and the ro-vibrational H$_2$ 1--0 S(1) emission line at $\lambda\approx2.12~\mu$m ({top-}right panel). We ntoat that for M82, 
  we only have access to H$_2$(4.69) as there is no observation available of H$_2$(2.12) emission 
  line in NIRSpec yet.  {The bottom row shows the same ratio as a function of Pfund($\lambda\approx3.29~\mu$m)$/$H$_2$($\lambda\approx4.69~\mu$m) on the left and of $G_0$ on the right panel. 
  For $G_0$, in purple, we indicate the area of AIB$_{\rm Irrad/Shielded}$ corresponding to galaxies.}
  For all panels, blue circles are the Orion Bar NIRSpec values for each spaxel. 
  Purple squares represent {the positions of the galaxies studied in these diagrams.} 
  {Spatially resolved galaxies} are represented with a red edge color and have error bars corresponding to 1~$\sigma$ percentiles (5$\%$ and 95$\%$ of fit values for all galaxy spaxels).
  For these points, the median values of all the fits are shown. 
  }
  \label{fig:ratio}
\end{figure*}

\section{Application to galaxies}
\label{section:galaxies}

\begin{table*}
\centering
\caption{Properties of galaxies with spectroscopy {in the 3.3~\um\ AIB range} obtained as part of the GOALS and TEMPLATES JWST-ERS programs.}
\label{table:galaxies}
\begin{tabular}{lcccccc}
\hline
Galaxy name    &  Redshift &I$_{\rm AIB}$/H$_2$(4.69) & I$_{\rm AIB}$/H$_2$(2.12)& Pfund(3.29)/H$_2$(4.69) & AIB$_{\rm Irrad/Shielded}$ & SFR\\
(1)   &  (2)   & (3) & (4) & (5) & (6) & (7) \\
\hline
IIZw96          & 0.0361$^1$ & 59.2 & 40.3 & 1.3  & 0.488 & 40--60$^1$\\
VV114           & 0.0202$^2$ & 81.3 & 54.6 & 0.73 & 0.411 & 38$^4$\\
NGC7469         & 0.01627$^5$& 135.1 & 73.5 & 0.97 & 1.067 & 10--30$^6$\\
NGC3256 nucl1   & 0.009364$^7$ & 217.4 & 89.3 & 2.6 & 0.689 & 30$^8$\\
NGC3256 nucl2   & 0.009364$^7$ & 69.4 & 32.7 & 2.5 & 0.412 & 50$^8$\\
STP0418-47      & 4.2248$^9$   & $\geq$64.9 & $\geq$59.2 & $\geq$5.8 & 0.075 & 350$^9$\\
M82 nucleus     & 0.00073$^{10}$ & 0.0061 & $\cdots$ & $\cdots$ &0.741 & 10$^{11}$\\
\hline
\end{tabular}\\
{
(1) Galaxy name;
(2) galaxy redshift;
(3) I$_{\rm AIB}$/H$_2$(4.69) = H$_2(\lambda 4.6946)/(\rm AIB_{3.2-3.7})$ flux ratio;
(4) I$_{\rm AIB}$/H$_2$(2.12) = H$_2(\lambda 2.1218)/(\rm AIB_{3.2-3.7})$ flux ratio;
(5) Pfund(3.29)/H$_2$(4.69) = Pfund($\lambda 3.297)/$H$_2(\lambda 4.6946)$ flux ratio;
(6) AIB$_{\rm Irrad/Shielded}$ = AIB$_{\rm Irrad}$/AIB$_{\rm Shielded}$;
(7) The galaxy SFR (in \mpy).
$^1$: \citet{GOALS_JWST_iizw96};
$^2$: \citet{GOALS_JWST_vv114};
$^3$: \citet{GOALS_JWST_vv114_b};
$^4$: \citet{VV114_SFR};
$^5$: \citet{GOALS_JWST_ngc7469};
$^6$: \citet{GOALS_JWST_ngc7469_b};
$^7$: \citet{Yuan_NGC3256};
$^8$: \citet{SakamotoK_14_ngc3256};
$^9$: \citet{SpilkerJ_23};
$^{10}$: \citet{MorabitoL_14_m82};
$^{11}$: \citet{Yoast_13_m82}.
}
\end{table*}

\subsection{Fitting global galaxy spectra using the templates}

In this section we now turn to galaxies observed with NIRSpec. 
We fit the 7 average spectra of these objects in our sample using a linear combination 
of emission lines, continuum and both AIB$_{\rm Irrad}$ and AIB$_{\rm Shielded}$ 
templates (we follow the same procedure as in Sect.~\ref{sec:spatial_dist_templates}
but accounting for galaxy redshift when fitting the spectra).
For galaxy SPT0418-47, as the spectrum has a relatively low SNR ($\approx 2$), 
emission lines are not detected and thus not included in the fit. 

\begin{figure*}[ht!]
  \centering
  \includegraphics[width=16cm]{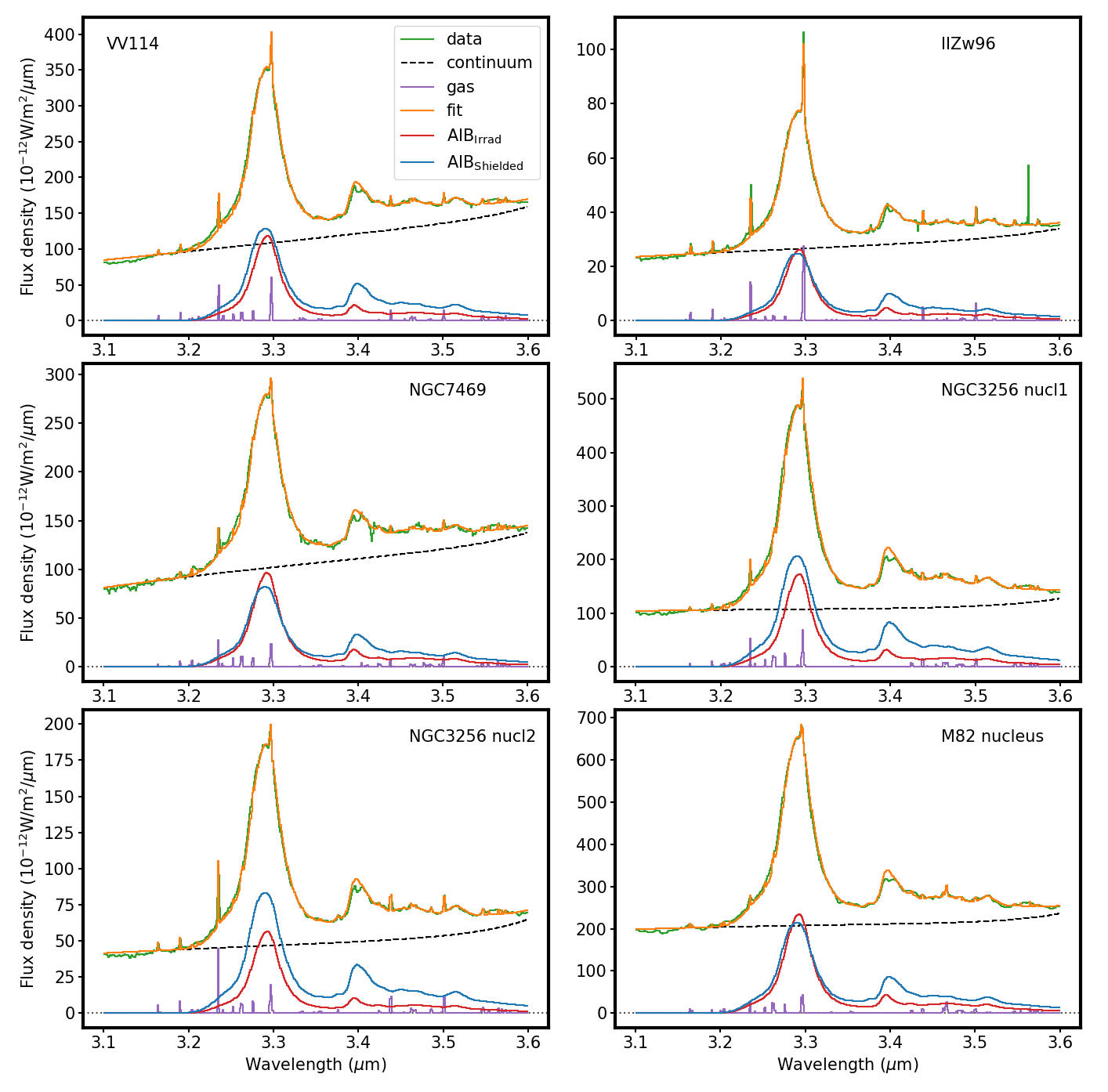}
  \caption{ ULIRGs from GOALS-JWST NIRSpec and M82 from NIRSpec MSA fitted spectra using a linear combination of continuum (dashed black lines), emission lines (purple solid lines) and the extracted  templates AIB$_{\rm Irrad}$ (in red) and AIB$_{\rm Shielded}$ (in blue). 
  Each panel corresponds to a galaxy whose name is indicated. 
  }
  \label{fig:galaxies}
\end{figure*}

{ Figure~\ref{fig:galaxies} shows all the fits obtained for nearby galaxies. 
The root mean square normalized error (see definition e.g. in \citet{Boubou_21}) 
is below 1\%. From those fits, we extract the AIB$_{\rm Irrad/Shielded}$ and 
compare them to the same observational parameters as for the Orion Bar 
(Fig.~\ref{fig:ratio}). 
For all galaxies in the study, the derived values overlap
with those of the Orion Bar. Interestingly, for all the nearby galaxies, the points are 
situated near the shoulder of the Orion Bar data.
This indicates that the AIB spectra from those nearby galaxies 
likely trace the irradiated surfaces of molecular clouds in the vicinity of massive stars. 
As can be seen in the bottom right panel of Fig.~\ref{fig:ratio}, 
the range of values of AIB$_{\rm Irrad/Shielded}$ derived for nearby 
galaxies is consistent with radiation field in the range of 
$G_0\sim 2-20\times10^3$ (Fig.~\ref{fig:ratio}).}

\begin{figure}[h!]
  \centering
  \includegraphics[width=9cm]{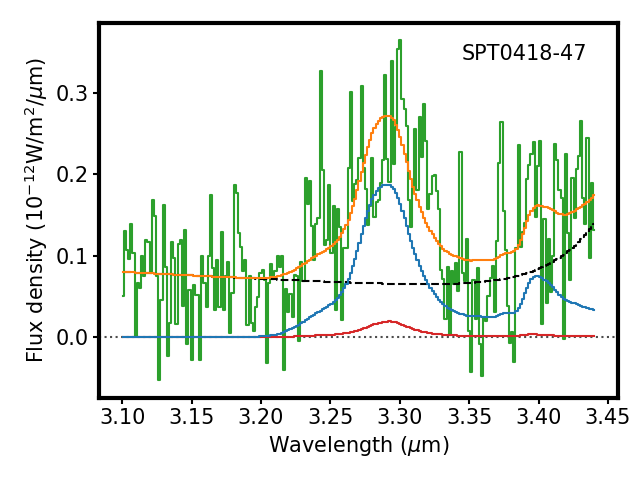}
  \caption{
  Same as Figure~\ref{fig:fit} for the SPT0418 $z=4.2$ galaxy using the extracted AIB$_{\rm Irrad}$ (fit shown in red) and AIB$_{\rm Shielded}$ (fit shown in blue) templates. 
  }
  \label{fig:fit_galaxy_high_z}
\end{figure}

\subsection{Fitting the spectra of spatially resolved galaxies}
\label{subsection:resolved_fitting}

In order to obtain complementary insights into the variations of the AIB emission within {local} galaxies,
we now perform a similar analysis {by} fitting the spectra at all spatial positions within the 
NIRSpec cubes of the GOALS program \footnote{For M82 however, this galaxy has no NIRSpec IFU observation 
available and we thus cannot perform a ``resolved'' fitting like the other five nearby galaxies}. 
The fitted spectra are extracted over an aperture of one pixel radius. 
We thus end up with approximately 1500 spectra per pointing.
Each spectrum is then fitted using the linear combination of the extracted templates (AIB$_{\rm Irrad}$ and AIB$_{\rm Shielded}$, 
linear and power-law continuum, and emission lines), as above. 
From these fits, we derive the maps of the AIB$_{\rm Irrad/Shielded}$ ratio for each galaxy. 
Those maps are shown in Figure~\ref{fig:galaxies_palpha}. 
On this figure, we also include the maps of the Paschen alpha line at 1.875~$\mu$m (left column),  
and of the integrated intensity of the 3.3~\um\ AIB. Emission in the Paschen alpha line is a tracer of 
Extreme UV ($E >13.6$ eV) photons, while the 3.3~\um\ integrated intensity is more sensitive to 
Far-UV ($E <13.6$ eV) photons. Thus, both are tracers of the radiative feedback from
OB stars on the ISM, yet with different physical conditions (respectively HII region vs PDR, 
\citealt{PeetersE_23}). {Since the evolution of the ratio of the 3.3 over 3.4~\um\ bands, and therefore the relative contribution of AIB$_{\rm Irrad}$ and AIB$_{\rm Shielded}$ depends on the FUV intensity \citep[][Fig.~\ref{fig:template_comparison} of this work]{joblin96}, we compare in Fig.~\ref{fig:galaxies_palpha}, the spatial distribution of AIB$_{\rm Irrad/Shielded}$ with that of the 3.3~\um\ band.}

In the case of IIZw96  AIB$_{\rm Irrad/Shielded}$ seems to follow the Paschen alpha and 
the 3.3~\um\ emissions at first look. However, looking in more detail,
the AIB$_{\rm Irrad/Shielded}$ ratio peaks near the central Paschen alpha and 3.3~\um\ knots but not necessarily
exactly at their positions. Also, some knots are seen peaking in Pa alpha and 3.3 emission
but not in AIB$_{\rm Irrad/Shielded}$. 
In VV114, AIB$_{\rm Irrad/Shielded}$ is bright in the North-East knot of 
3.3 \um\ and Paschen alpha emission, but is small near the nucleus where 
both Paschen alpha and 3.3 \um\ emission peak.  
In NGC 7469, AIB$_{\rm Irrad/Shielded}$ follows the 
knots of active star formation, however the distributions
are no fully co-spatial. Instead, it appears that 
AIB$_{\rm Irrad/Shielded}$ is somewhat on the edges of these knots and seems to miss the star-forming region in the South. 
The situation near the nucleus is more difficult to assess 
due to saturation of the NIRSpec spectra. 
For NGC 3256 nucleus 1, AIB$_{\rm Irrad/Shielded}$ peaks close to 
the peak of 3.3~\um\ and Paschen Alpha emission near the nucleus of the
galaxy. However, the spatial pattern differs: AIB$_{\rm Irrad/Shielded}$ 
shows a more diffuse emission, linked to the starforming knots,
but not completely co-located, as in NGC 7469. 
Finally, in NGC 3256 nucleus 2, the distribution of AIB$_{\rm Irrad/Shielded}$
seems completely unrelated to Paschen alpha or the 3.3~\um\ emission.
We do see however --as in all other galaxies, a peak of AIB$_{\rm Irrad/Shielded}$
near a starforming knot at the Northwest of the field of view  (Fig.~\ref{fig:galaxies_palpha} ).
Comparing the AIB$_{\rm Irrad/Shielded}$ to the HST visible images, 
it appears that low values of AIB$_{\rm Irrad/Shielded}$ seem to correspond to dark regions 
of the HST images (Fig.~\ref{fig:galaxies_palpha}), where visual extinction is high, i.e. regions 
with large amounts of molecular gas. 
In summary, AIB$_{\rm Irrad/Shielded}$ does seem to be related to the radiative feedback from massive
stars in the observed galaxies. As in the Orion Bar, this ratio is high in the vicinity of massive OB 
stars, and is much lower in regions with higher extinction and larger molecular gas fractions. 
This is also compatible with the findings of \citet{LaiT_20} who found a correlation between 
the 3.4/3.3 \um\ band ratio and the radiation field intensity, suggesting that aliphatic (3.4) 
component is more prone to photo-destruction under intense radiation environment.

\begin{figure*}[]
  \centering
  \includegraphics[width=16cm]{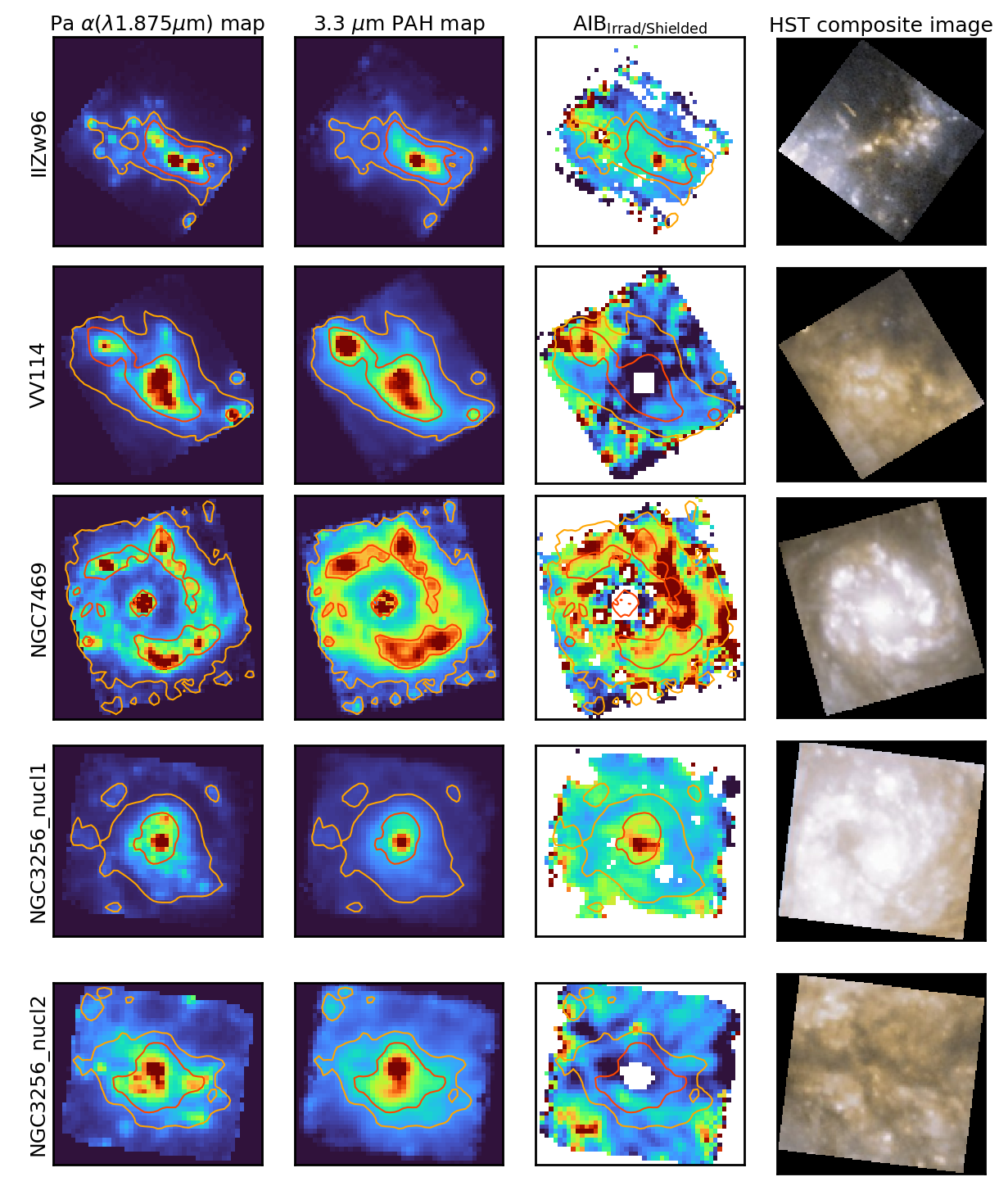}
  \caption{
  \textit{First column: }NIRSpec continuum-subtracted Paschen $\alpha$ emission maps.
  \textit{Second column: }NIRSpec continuum-subtracted 3.3~$\mu$m emission maps with intensity contours. The contours are shown on each panel for each galaxy for better comparison. 
  \textit{Third column: }AIB$_{\rm Irrad/Shielded}$ contribution maps for each galaxy. 
  On this column, saturated spaxels are represented in white. Note that we do not take into account edge spaxels as they are too noisy and mostly contain artefacts.
  \textit{Last column: }Hubble Space Telescope colored composite images (F435W and F814W filters) of each galaxy.
  }
  \label{fig:galaxies_palpha}
\end{figure*}

\subsection{Fitting a high redshift AIB spectrum}

In the last step of our analysis we fit the AIB spectrum of the $z=4.2$
galaxy SPT0418 obtained with MIRI using our template spectra. The result
of this fit is shown in Figure~\ref{fig:fit_galaxy_high_z}. 
{Although the SNR is poor on this observation, this suggests that our extracted templates could be applied to a high redshift galaxy.}
For this source, the H$_2$ lines are not detected, thus we derive an upper limit
for the ratios {between the H$_2$ line intensity and $I_{\rm AIB}$} using the measured RMS noise at the wavelengths
of the 2.12 and 4.69 \um\ H$_2$ lines. We then place SPT0418 in the diagrams of 
Fig.~\ref{fig:ratio}. Interestingly, it appears to fall in a different region: 
while the H$_2$ over AIB ratio is only an upper limit,
the AIB$_{\rm Irrad/Shielded}$ is clearly lower than for nearby galaxies. 
This lower value of AIB$_{\rm Irrad/Shielded}$ with respect to galaxies of the 
local universe would indicate that less highly irradiated gas is present in 
SPT0418. This result is quite counter intuitive considering the large 
SFR of this galaxy. However, given the noisy nature of the spectrum (Fig.~\ref{fig:fit_galaxy_high_z}) 
the reliability of the derived value of AIB$_{\rm Irrad/Shielded}$ needs to be confirmed with 
deeper observations of this source with MIRI.

\section{Conclusion}
\label{section:conclusion}

In this paper, using JWST NIRSpec observations of the Orion Bar, we extracted two template spectra
AIB$_{\rm Irrad}$ and AIB$_{\rm Shielded}$ {which give a good representation of} the AIB emission from this PDR in the 3.2 to 3.6 \um\ range. 
We have shown that the ratio between these two templates, AIB$_{\rm Irrad / Shielded}$,  
is linked to the physical conditions in the PDR, with high values corresponding to regions where the UV field is strongest and where H$_2$ is dissociated and low values corresponding to regions where the UV field is weakest and where H$_2$ can form.
{We fitted NIRSPec spectra of nearby galaxies using these two templates} 
and extracted maps of the 
AIB$_{\rm Irrad / Shielded}$ ratio. We showed that this ratio follows the 
radiative feedback from massive stars. Finally, {we used our templates to fit} the spectrum of a high redshift galaxy, showing the potential of such an approach to derive physical conditions 
of the ISM in these objects. 
These templates could also be used, for instance, to fit the spectral energy distribution 
of galaxies, in order to constrain the intrinsic properties of galaxies {such as} their initial mass function distribution or even their SFR history. 
Additional spectroscopic observations  with JWST covering the range of the 3.3 \um\ AIB for both low and high redshift (up to $z\approx 7$) galaxies are needed {to confirm and refine the potential of the approach presented here to determine the global physical conditions of the ISM in galaxies.}

\section*{Acknowledgments}
{We thank the referee for her/his comments and suggestions which helped to improve the clarity of the manuscript.}
This work is based [in part] on observations made with the NASA/ESA/CSA James Webb Space Telescope. The data were obtained from the Mikulski Archive for Space Telescopes at the Space Telescope Science Institute, which is operated by the Association of Universities for Research in Astronomy, Inc., under NASA contract NAS 5-03127 for JWST. These observations are associated with programs \#1288, \#1328, \#1355 and \#2677.
IS, OB, AC are funded by the Centre National d'Etudes Spatiales (CNES) through the APR program. 
EP acknowledges support from the University of Western Ontario, the Institute for Earth and Space Exploration, the Canadian Space Agency (CSA, 22JWGO1-16), and the Natural Sciences and Engineering Research Council of Canada. 

\bibliography{references}{}

\begin{thebibliography}{46}
\expandafter\ifx\csname natexlab\endcsname\relax\def\natexlab#1{#1}\fi

\bibitem[{{Allamandola} {et~al.}(1989){Allamandola}, {Tielens}, \&
  {Barker}}]{AllamandolaL_89}
{Allamandola}, L.~J., {Tielens}, A.~G.~G.~M., \& {Barker}, J.~R. 1989, \apjs,
  71, 733

\bibitem[{{Armus} {et~al.}(2023){Armus}, {Lai}, {U}, {Larson}, {Diaz-Santos},
  {Evans}, {Malkan}, {Rich}, {Medling}, {Law}, {Inami}, {Muller-Sanchez},
  {Charmandaris}, {van der Werf}, {Stierwalt}, {Linden}, {Privon},
  {Barcos-Mu{\~n}oz}, {Hayward}, {Song}, {Appleton}, {Aalto}, {Bohn},
  {B{\"o}ker}, {Brown}, {Finnerty}, {Howell}, {Iwasawa}, {Kemper}, {Marshall},
  {Mazzarella}, {McKinney}, {Murphy}, {Sanders}, \&
  {Surace}}]{GOALS_JWST_ngc7469_e}
{Armus}, L., {Lai}, T., {U}, V., {et~al.} 2023, \apjl, 942, L37

\bibitem[{{Bakes} \& {Tielens}(1994)}]{BakesE_94}
{Bakes}, E.~L.~O. \& {Tielens}, A.~G.~G.~M. 1994, \apj, 427, 822

\bibitem[{{Basu-Zych} {et~al.}(2013){Basu-Zych}, {Lehmer}, {Hornschemeier},
  {Gon{\c{c}}alves}, {Fragos}, {Heckman}, {Overzier}, {Ptak}, \&
  {Schiminovich}}]{VV114_SFR}
{Basu-Zych}, A.~R., {Lehmer}, B.~D., {Hornschemeier}, A.~E., {et~al.} 2013,
  \apj, 774, 152

\bibitem[{{Bern{\'e}} {et~al.}(2022{\natexlab{a}}){Bern{\'e}}, {Foschino},
  {Jalabert}, \& {Joblin}}]{BerneO_22}
{Bern{\'e}}, O., {Foschino}, S., {Jalabert}, F., \& {Joblin}, C.
  2022{\natexlab{a}}, \aap, 667, A159

\bibitem[{{Bern{\'e}} {et~al.}(2022{\natexlab{b}}){Bern{\'e}}, {Habart},
  {Peeters}, {Abergel}, {Bergin}, {Bernard-Salas}, {Bron}, {Cami}, {Dartois},
  {Fuente}, {Goicoechea}, {Gordon}, {Okada}, {Onaka}, {Robberto}, {R{\"o}llig},
  {Tielens}, {Vicente}, {Wolfire}, {Alarc{\'o}n}, {Boersma}, {Canin}, {Chown},
  {Dicken}, {Languignon}, {Le Gal}, {Pound}, {Trahin}, {Simmer}, {Sidhu}, {Van
  De Putte}, {Cuadrado}, {Guilloteau}, {Maragkoudakis}, {Schefter}, {Schirmer},
  {Cazaux}, {Aleman}, {Allamandola}, {Auchettl}, {Baratta}, {Bejaoui}, {Bera},
  {Bilalbegovi{\'c}}, {Black}, {Boulanger}, {Bouwman}, {Brandl}, {Brechignac},
  {Br{\"u}nken}, {Burkhardt}, {Candian}, {Cernicharo}, {Chabot}, {Chakraborty},
  {Champion}, {Colgan}, {Cooke}, {Coutens}, {Cox}, {Demyk}, {Donovan Meyer},
  {Engrand}, {Foschino}, {Garc{\'\i}a-Lario}, {Gavilan}, {Gerin}, {Godard},
  {Gottlieb}, {Guillard}, {Gusdorf}, {Hartigan}, {He}, {Herbst}, {Hornekaer},
  {J{\"a}ger}, {Janot-Pacheco}, {Joblin}, {Kaufman}, {Kemper}, {Kendrew},
  {Kirsanova}, {Klaassen}, {Knight}, {Kwok}, {Labiano}, {Lai}, {Lee},
  {Lefloch}, {Le Petit}, {Li}, {Linz}, {Mackie}, {Madden}, {Mascetti},
  {McGuire}, {Merino}, {Micelotta}, {Misselt}, {Morse}, {Mulas}, {Neelamkodan},
  {Ohsawa}, {Omont}, {Paladini}, {Palumbo}, {Pathak}, {Pendleton},
  {Petrignani}, {Pino}, {Puga}, {Rangwala}, {Rapacioli}, {Ricca},
  {Roman-Duval}, {Roser}, {Roueff}, {Rouill{\'e}}, {Salama}, {Sales},
  {Sandstrom}, {Sarre}, {Sciamma-O'Brien}, {Sellgren}, {Shannon}, {Shenoy},
  {Teyssier}, {Thomas}, {Togi}, {Verstraete}, {Witt}, {Wootten}, {Ysard},
  {Zettergren}, {Zhang}, {Zhang}, \& {Zhen}}]{PDRs4All_22}
{Bern{\'e}}, O., {Habart}, {\'E}., {Peeters}, E., {et~al.} 2022{\natexlab{b}},
  \pasp, 134, 054301

\bibitem[{{Bern{\'e}} {et~al.}(2007){Bern{\'e}}, {Joblin}, {Deville}, {Smith},
  {Rapacioli}, {Bernard}, {Thomas}, {Reach}, \& {Abergel}}]{Berne_O_07}
{Bern{\'e}}, O., {Joblin}, C., {Deville}, Y., {et~al.} 2007, \aap, 469, 575

\bibitem[{Bernstein {et~al.}(1996)Bernstein, Sandford, \&
  Allamandola}]{bernstein1996hydrogenated}
Bernstein, M.~P., Sandford, S.~A., \& Allamandola, L.~J. 1996, The
  Astrophysical Journal, 472, L127

\bibitem[{{Bianchin} {et~al.}(2023){Bianchin}, {U}, {Song}, {Lai}, {Remigio},
  {Barcos-Munoz}, {Diaz-Santos}, {Armus}, {Inami}, {Larson}, {Evans}, {Boker},
  {Kader}, {Linden}, {Charmandaris}, {Malkan}, {Rich}, {Bohn}, {Medling},
  {Stierwalt}, {Mazzarella}, {Law}, {Privon}, {Aalto}, {Appleton}, {Brown},
  {Buiten}, {Finnerty}, {Hayward}, {Howell}, {Iwasawa}, {Kemper}, {Marshall},
  {McKinney}, {Muller-Sanchez}, {Murphy}, {van der Werf}, {Sanders}, \&
  {Surace}}]{BianchinM_23}
{Bianchin}, M., {U}, V., {Song}, Y., {et~al.} 2023, arXiv e-prints,
  arXiv:2308.00209

\bibitem[{{Bohn} {et~al.}(2023){Bohn}, {Inami}, {Diaz-Santos}, {Armus},
  {Linden}, {U}, {Surace}, {Larson}, {Evans}, {Hoshioka}, {Lai}, {Song},
  {Mazzarella}, {Barcos-Munoz}, {Charmandaris}, {Howell}, {Medling}, {Privon},
  {Rich}, {Stierwalt}, {Aalto}, {B{\"o}ker}, {Brown}, {Iwasawa}, {Malkan}, {van
  der Werf}, {Appleton}, {Hayward}, {Kemper}, {Law}, {Marshall}, {Murphy}, \&
  {Sanders}}]{GOALS_JWST_ngc7469_c}
{Bohn}, T., {Inami}, H., {Diaz-Santos}, T., {et~al.} 2023, \apjl, 942, L36

\bibitem[{{Boulais} {et~al.}(2021){Boulais}, {Bern{\'e}}, {Faury}, \&
  {Deville}}]{Boubou_21}
{Boulais}, A., {Bern{\'e}}, O., {Faury}, G., \& {Deville}, Y. 2021, \aap, 647,
  A105

\bibitem[{{Chown} {et~al.}(2023){Chown}, {Sidhu}, {Peeters}, {Tielens}, {Cami},
  {Bern{\'e}}, {Habart}, {Alarc{\'o}n}, {Canin}, {Schroetter}, {Trahin}, {Van
  De Putte}, {Abergel}, {Bergin}, {Bernard-Salas}, {Boersma}, {Bron},
  {Cuadrado}, {Dartois}, {Dicken}, {El-Yajouri}, {Fuente}, {Goicoechea},
  {Gordon}, {Issa}, {Joblin}, {Kannavou}, {Khan}, {Lacinbala}, {Languignon},
  {Le Gal}, {Maragkoudakis}, {Meshaka}, {Okada}, {Onaka}, {Pasquini}, {Pound},
  {Robberto}, {R{\"o}llig}, {Schefter}, {Schirmer}, {Vicente}, {Wolfire},
  {Zannese}, {Aleman}, {Allamandola}, {Auchettl}, {Baratta}, {Bejaoui}, {Bera},
  {Black}, {Boulanger}, {Bouwman}, {Brandl}, {Brechignac}, {Br{\"u}nken},
  {Buragohain}, {Burkhardt}, {Candian}, {Cazaux}, {Cernicharo}, {Chabot},
  {Chakraborty}, {Champion}, {Colgan}, {Cooke}, {Coutens}, {Cox}, {Demyk},
  {Donovan Meyer}, {Foschino}, {Garc{\'\i}a-Lario}, {Gavilan}, {Gerin},
  {Gottlieb}, {Guillard}, {Gusdorf}, {Hartigan}, {He}, {Herbst}, {Hornekaer},
  {J{\"a}ger}, {Janot-Pacheco}, {Kaufman}, {Kemper}, {Kendrew}, {Kirsanova},
  {Klaassen}, {Kwok}, {Labiano}, {Lai}, {Lee}, {Lefloch}, {Le Petit}, {Li},
  {Linz}, {Mackie}, {Madden}, {Mascetti}, {McGuire}, {Merino}, {Micelotta},
  {Misselt}, {Morse}, {Mulas}, {Neelamkodan}, {Ohsawa}, {Omont}, {Paladini},
  {Palumbo}, {Pathak}, {Pendleton}, {Petrignani}, {Pino}, {Puga}, {Rangwala},
  {Rapacioli}, {Ricca}, {Roman-Duval}, {Roser}, {Roueff}, {Rouill{\'e}},
  {Salama}, {Sales}, {Sandstrom}, {Sarre}, {Sciamma-O'Brien}, {Sellgren},
  {Shenoy}, {Teyssier}, {Thomas}, {Togi}, {Verstraete}, {Witt}, {Wootten},
  {Zettergren}, {Zhang}, {Zhang}, \& {Zhen}}]{ChownR_embarrassment23}
{Chown}, R., {Sidhu}, A., {Peeters}, E., {et~al.} 2023, arXiv e-prints,
  arXiv:2308.16733

\bibitem[{{Draine} {et~al.}(2007){Draine}, {Dale}, {Bendo}, {Gordon}, {Smith},
  {Armus}, {Engelbracht}, {Helou}, {Kennicutt}, {Li}, {Roussel}, {Walter},
  {Calzetti}, {Moustakas}, {Murphy}, {Rieke}, {Bot}, {Hollenbach}, {Sheth}, \&
  {Teplitz}}]{DraineB_07}
{Draine}, B.~T., {Dale}, D.~A., {Bendo}, G., {et~al.} 2007, \apj, 663, 866

\bibitem[{{Evans} {et~al.}(2022){Evans}, {Frayer}, {Charmandaris}, {Armus},
  {Inami}, {Surace}, {Linden}, {Soifer}, {Diaz-Santos}, {Larson}, {Rich},
  {Song}, {Barcos-Munoz}, {Mazzarella}, {Privon}, {U}, {Medling}, {B{\"o}ker},
  {Aalto}, {Iwasawa}, {Howell}, {van der Werf}, {Appleton}, {Bohn}, {Brown},
  {Hayward}, {Hoshioka}, {Kemper}, {Lai}, {Law}, {Malkan}, {Marshall},
  {Murphy}, {Sanders}, \& {Stierwalt}}]{GOALS_JWST_vv114}
{Evans}, A.~S., {Frayer}, D.~T., {Charmandaris}, V., {et~al.} 2022, \apjl, 940,
  L8

\bibitem[{{Foschino} {et~al.}(2019){Foschino}, {Bern{\'e}}, \&
  {Joblin}}]{Foscino_19}
{Foschino}, S., {Bern{\'e}}, O., \& {Joblin}, C. 2019, \aap, 632, A84

\bibitem[{{Habart} {et~al.}(2023){Habart}, {Peeters}, {Bern{\'e}}, {Trahin},
  {Canin}, {Chown}, {Sidhu}, {Van De Putte}, {Alarc{\'o}n}, {Schroetter},
  {Dartois}, {Vicente}, {Abergel}, {Bergin}, {Bernard-Salas}, {Boersma},
  {Bron}, {Cami}, {Cuadrado}, {Dicken}, {Elyajouri}, {Fuente}, {Goicoechea},
  {Gordon}, {Issa}, {Joblin}, {Kannavou}, {Khan}, {Lacinbala}, {Languignon},
  {Le Gal}, {Maragkoudakis}, {Meshaka}, {Okada}, {Onaka}, {Pasquini}, {Pound},
  {Robberto}, {R{\"o}llig}, {Schefter}, {Schirmer}, {Tabone}, {Tielens},
  {Wolfire}, {Zannese}, {Ysard}, {Miville-Deschenes}, {Aleman}, {Allamandola},
  {Auchettl}, {Baratta}, {Bejaoui}, {Bera}, {Black}, {Boulanger}, {Bouwman},
  {Brandl}, {Brechignac}, {Br{\"u}nken}, {Buragohain}, {Burkhardt}, {Candian},
  {Cazaux}, {Cernicharo}, {Chabot}, {Chakraborty}, {Champion}, {Colgan},
  {Cooke}, {Coutens}, {Cox}, {Demyk}, {Donovan Meyer}, {Foschino},
  {Garc{\'\i}a-Lario}, {Gavilan}, {Gerin}, {Gottlieb}, {Guillard}, {Gusdorf},
  {Hartigan}, {He}, {Herbst}, {Hornekaer}, {J{\"a}ger}, {Janot-Pacheco},
  {Kaufman}, {Kemper}, {Kendrew}, {Kirsanova}, {Klaassen}, {Kwok}, {Labiano},
  {Lai}, {Lee}, {Lefloch}, {Le Petit}, {Li}, {Linz}, {Mackie}, {Madden},
  {Mascetti}, {McGuire}, {Merino}, {Micelotta}, {Misselt}, {Morse}, {Mulas},
  {Neelamkodan}, {Ohsawa}, {Omont}, {Paladini}, {Palumbo}, {Pathak},
  {Pendleton}, {Petrignani}, {Pino}, {Puga}, {Rangwala}, {Rapacioli}, {Ricca},
  {Roman-Duval}, {Roser}, {Roueff}, {Rouill{\'e}}, {Salama}, {Sales},
  {Sandstrom}, {Sarre}, {Sciamma-O'Brien}, {Sellgren}, {Shenoy}, {Teyssier},
  {Thomas}, {Togi}, {Verstraete}, {Witt}, {Wootten}, {Zettergren}, {Zhang},
  {Zhang}, \& {Zhen}}]{HabartE_imaging23}
{Habart}, E., {Peeters}, E., {Bern{\'e}}, O., {et~al.} 2023, arXiv e-prints,
  arXiv:2308.16732

\bibitem[{{Habing}(1968)}]{Habing68}
{Habing}, H.~J. 1968, 19, 421

\bibitem[{{Inami} {et~al.}(2022){Inami}, {Surace}, {Armus}, {Evans}, {Larson},
  {Barcos-Munoz}, {Stierwalt}, {Mazzarella}, {Privon}, {Song}, {Linden},
  {Hayward}, {B{\"o}ker}, {U}, {Bohn}, {Charmandaris}, {Diaz-Santos}, {Howell},
  {Lai}, {Medling}, {Rich}, {Aalto}, {Appleton}, {Brown}, {Hoshioka},
  {Iwasawa}, {Kemper}, {Law}, {Malkan}, {Marshall}, {Murphy}, {Sanders}, \&
  {van der Werf}}]{GOALS_JWST_iizw96}
{Inami}, H., {Surace}, J., {Armus}, L., {et~al.} 2022, \apjl, 940, L6

\bibitem[{{Joblin} {et~al.}(1996){Joblin}, {Tielens}, {Allamandola}, \&
  {Geballe}}]{joblin96}
{Joblin}, C., {Tielens}, A.~G.~G.~M., {Allamandola}, L.~J., \& {Geballe}, T.~R.
  1996, \apj, 458, 610

\bibitem[{{Kim} {et~al.}(2012){Kim}, {Im}, {Lee}, {Lee}, {Jun}, {Nakagawa},
  {Matsuhara}, {Wada}, {Oyabu}, {Takagi}, {Inami}, {Ohyama}, {Yamada}, {Helou},
  {Armus}, \& {Shi}}]{KimJ_12}
{Kim}, J.~H., {Im}, M., {Lee}, H.~M., {et~al.} 2012, \apj, 760, 120

\bibitem[{{Lai} {et~al.}(2023){Lai}, {Armus}, {Bianchin}, {D{\'\i}az-Santos},
  {Linden}, {Privon}, {Inami}, {U}, {Bohn}, {Evans}, {Larson}, {Hensley},
  {Smith}, {Malkan}, {Song}, {Stierwalt}, {van der Werf}, {McKinney}, {Aalto},
  {Buiten}, {Rich}, {Charmandaris}, {Appleton}, {Barcos-Mu{\~n}oz},
  {B{\"o}ker}, {Finnerty}, {Kader}, {Law}, {Medling}, {Brown}, {Hayward},
  {Howell}, {Iwasawa}, {Kemper}, {Marshall}, {Mazzarella},
  {M{\"u}ller-S{\'a}nchez}, {Murphy}, {Sanders}, \& {Surace}}]{LaiT_23}
{Lai}, T. S.~Y., {Armus}, L., {Bianchin}, M., {et~al.} 2023, \apjl, 957, L26

\bibitem[{{Lai} {et~al.}(2022){Lai}, {Armus}, {U}, {D{\'\i}az-Santos},
  {Larson}, {Evans}, {Malkan}, {Appleton}, {Rich}, {M{\"u}ller-S{\'a}nchez},
  {Inami}, {Bohn}, {McKinney}, {Finnerty}, {Law}, {Linden}, {Medling},
  {Privon}, {Song}, {Stierwalt}, {van der Werf}, {Barcos-Mu{\~n}oz}, {Smith},
  {Togi}, {Aalto}, {B{\"o}ker}, {Charmandaris}, {Howell}, {Iwasawa}, {Kemper},
  {Mazzarella}, {Murphy}, {Brown}, {Hayward}, {Marshall}, {Sanders}, \&
  {Surace}}]{GOALS_JWST_ngc7469_b}
{Lai}, T. S.~Y., {Armus}, L., {U}, V., {et~al.} 2022, \apjl, 941, L36

\bibitem[{{Lai} {et~al.}(2020){Lai}, {Smith}, {Baba}, {Spoon}, \&
  {Imanishi}}]{LaiT_20}
{Lai}, T. S.~Y., {Smith}, J.~D.~T., {Baba}, S., {Spoon}, H. W.~W., \&
  {Imanishi}, M. 2020, \apj, 905, 55

\bibitem[{{Lee} \& {Seung}(1999)}]{LeeD_nmf}
{Lee}, D.~D. \& {Seung}, H.~S. 1999, \nat, 401, 788

\bibitem[{{Li}(2020)}]{LiA_20}
{Li}, A. 2020, Nature Astronomy, 4, 339

\bibitem[{{Linden} {et~al.}(2023){Linden}, {Evans}, {Armus}, {Rich}, {Larson},
  {Lai}, {Privon}, {U}, {Inami}, {Bohn}, {Song}, {Barcos-Mu{\~n}oz},
  {Charmandaris}, {Medling}, {Stierwalt}, {Diaz-Santos}, {B{\"o}ker}, {van der
  Werf}, {Aalto}, {Appleton}, {Brown}, {Hayward}, {Howell}, {Iwasawa},
  {Kemper}, {Frayer}, {Law}, {Malkan}, {Marshall}, {Mazzarella}, {Murphy},
  {Sanders}, \& {Surace}}]{GOALS_JWST_vv114_c}
{Linden}, S.~T., {Evans}, A.~S., {Armus}, L., {et~al.} 2023, \apjl, 944, L55

\bibitem[{{Marciniak} {et~al.}(2021){Marciniak}, {Joblin}, {Mulas},
  {Mundlapati}, \& {Bonnamy}}]{Marciniak2021}
{Marciniak}, A., {Joblin}, C., {Mulas}, G., {Mundlapati}, V.~R., \& {Bonnamy},
  A. 2021, \aap, 652, A42

\bibitem[{{McKinney} {et~al.}(2020){McKinney}, {Pope}, {Armus}, {Chary},
  {D{\'\i}az-Santos}, {Dickinson}, \& {Kirkpatrick}}]{McKinneyJ_20}
{McKinney}, J., {Pope}, A., {Armus}, L., {et~al.} 2020, \apj, 892, 119

\bibitem[{{Morabito} {et~al.}(2014){Morabito}, {Oonk}, {Salgado}, {Toribio},
  {R{\"o}ttgering}, {Tielens}, {Beck}, {Adebahr}, {Best}, {Beswick},
  {Bonafede}, {Brunetti}, {Br{\"u}ggen}, {Chy{\.z}y}, {Conway}, {van Driel},
  {Gregson}, {Haverkorn}, {Heald}, {Horellou}, {Horneffer}, {Iacobelli},
  {Jarvis}, {Marti-Vidal}, {Miley}, {Mulcahy}, {Orr{\'u}}, {Pizzo}, {Scaife},
  {Varenius}, {van Weeren}, {White}, \& {Wise}}]{MorabitoL_14_m82}
{Morabito}, L.~K., {Oonk}, J.~B.~R., {Salgado}, F., {et~al.} 2014, \apjl, 795,
  L33

\bibitem[{{Peeters} {et~al.}(2023){Peeters}, {Habart}, {Berne}, {Sidhu},
  {Chown}, {Van De Putte}, {Trahin}, {Schroetter}, {Canin}, {Alarcon},
  {Schefter}, {Khan}, {Pasquini}, {Tielens}, {Wolfire}, {Dartois},
  {Goicoechea}, {Maragkoudakis}, {Onaka}, {Pound}, {Vicente}, {Abergel},
  {Bergin}, {Bernard-Salas}, {Boersma}, {Bron}, {Cami}, {Cuadrado}, {Dicken},
  {Elyajour}, {Fuente}, {Gordon}, {Issa}, {Joblin}, {Kannavou}, {Lacinbala},
  {Languignon}, {Le Gal}, {Meshaka}, {Okada}, {Robberto}, {Roellig},
  {Schirmer}, {Tabone}, {Zannese}, {Aleman}, {Allamandola}, {Auchettl},
  {Baratta}, {Bejaoui}, {Bera}, {Black}, {Boulanger}, {Bouwman}, {Brandl},
  {Brechignac}, {Brunken}, {Buragohain}, {Burkhardt}, {Candian}, {Cazaux},
  {Cernicharo}, {Chabot}, {Chakraborty}, {Champion}, {Colgan}, {Cooke},
  {Coutens}, {Cox}, {Demyk}, {Donovan Meyer}, {Foschino}, {Garcia-Lario},
  {Gerin}, {Gottlieb}, {Guillard}, {Gusdorf}, {Hartigan}, {He}, {Herbst},
  {Hornekaer}, {Jager}, {Janot-Pacheco}, {Kaufman}, {Kendrew}, {Kirsanova},
  {Klaassen}, {Kwok}, {Labiano}, {Lai}, {Lee}, {Lefloch}, {Le Petit}, {Li},
  {Linz}, {Mackie}, {Madden}, {Mascetti}, {McGuire}, {Merino}, {Micelotta},
  {Misselt}, {Morse}, {Mulas}, {Neelamkodan}, {Ohsawa}, {Paladini}, {Palumbo},
  {Pathak}, {Pendleton}, {Petrignani}, {Pino}, {Puga}, {Rangwala}, {Rapacioli},
  {Ricca}, {Roman-Duval}, {Roser}, {Roueff}, {Rouille}, {Salama}, {Sales},
  {Sandstrom}, {Sarre}, {Sciamma-O'Brien}, {Sellgren}, {Shenoy}, {Teyssier},
  {Thomas}, {Togi}, {Verstraete}, {Witt}, {Wootten}, {Ysard}, {Zettergren},
  {Zhang}, {Zhang}, \& {Zhen}}]{PeetersE_23}
{Peeters}, E., {Habart}, E., {Berne}, O., {et~al.} 2023, arXiv e-prints,
  arXiv:2310.08720

\bibitem[{{Pilleri} {et~al.}(2015){Pilleri}, {Joblin}, {Boulanger}, \&
  {Onaka}}]{Pilleri_15}
{Pilleri}, P., {Joblin}, C., {Boulanger}, F., \& {Onaka}, T. 2015, \aap, 577,
  A16

\bibitem[{{Rich} {et~al.}(2023){Rich}, {Aalto}, {Evans}, {Charmandaris},
  {Privon}, {Lai}, {Inami}, {Linden}, {Armus}, {Diaz-Santos}, {Appleton},
  {Barcos-Mu{\~n}oz}, {B{\"o}ker}, {Larson}, {Law}, {Malkan}, {Medling},
  {Song}, {U}, {van der Werf}, {Bohn}, {Brown}, {Finnerty}, {Hayward},
  {Howell}, {Iwasawa}, {Kemper}, {Marshall}, {Mazzarella}, {McKinney},
  {Muller-Sanchez}, {Murphy}, {Sanders}, {Soifer}, {Stierwalt}, \&
  {Surace}}]{GOALS_JWST_vv114_b}
{Rich}, J., {Aalto}, S., {Evans}, A.~S., {et~al.} 2023, \apjl, 944, L50

\bibitem[{{Riechers} {et~al.}(2014){Riechers}, {Pope}, {Daddi}, {Armus},
  {Carilli}, {Walter}, {Hodge}, {Chary}, {Morrison}, {Dickinson},
  {Dannerbauer}, \& {Elbaz}}]{RiechersD_14}
{Riechers}, D.~A., {Pope}, A., {Daddi}, E., {et~al.} 2014, \apj, 786, 31

\bibitem[{{Rigopoulou} {et~al.}(2021){Rigopoulou}, {Barale}, {Clary}, {Shan},
  {Alonso-Herrero}, {Garc{\'\i}a-Bernete}, {Hunt}, {Kerkeni},
  {Pereira-Santaella}, \& {Roche}}]{RigopoulouD_21}
{Rigopoulou}, D., {Barale}, M., {Clary}, D.~C., {et~al.} 2021, \mnras, 504,
  5287

\bibitem[{{Sakamoto} {et~al.}(2014){Sakamoto}, {Aalto}, {Combes}, {Evans}, \&
  {Peck}}]{SakamotoK_14_ngc3256}
{Sakamoto}, K., {Aalto}, S., {Combes}, F., {Evans}, A., \& {Peck}, A. 2014,
  \apj, 797, 90

\bibitem[{{Sandstrom} {et~al.}(2023){Sandstrom}, {Chastenet}, {Sutter},
  {Leroy}, {Egorov}, {Williams}, {Bolatto}, {Boquien}, {Cao}, {Dale}, {Lee},
  {Rosolowsky}, {Schinnerer}, {Barnes}, {Belfiore}, {Bigiel}, {Chevance},
  {Grasha}, {Groves}, {Hassani}, {Hughes}, {Klessen}, {Kruijssen}, {Larson},
  {Liu}, {Lopez}, {Meidt}, {Murphy}, {Sormani}, {Thilker}, \&
  {Watkins}}]{SandstromK_23}
{Sandstrom}, K., {Chastenet}, J., {Sutter}, J., {et~al.} 2023, arXiv e-prints,
  arXiv:2301.00854

\bibitem[{{Spilker} {et~al.}(2023){Spilker}, {Phadke}, {Aravena}, {Archipley},
  {Bayliss}, {Birkin}, {Bethermin}, {Burgoyne}, {Cathey}, {Chapman}, {Dahle},
  {Gonzalez}, {Gururajan}, {Hayward}, {Hezaveh}, {Hill}, {Hutchison}, {Kim},
  {Kim}, {Law}, {Legin}, {Malkan}, {Marrone}, {Murphy}, {Narayanan}, {Navarre},
  {Olivier}, {Rich}, {Rigby}, {Reuter}, {Rhoads}, {Sharon}, {Smith},
  {Solimano}, {Sulzenauer}, {Vieira}, {Weiss}, \& {Whitaker}}]{SpilkerJ_23}
{Spilker}, J.~S., {Phadke}, K.~A., {Aravena}, M., {et~al.} 2023, arXiv
  e-prints, arXiv:2306.03152

\bibitem[{{U} {et~al.}(2022){U}, {Lai}, {Bianchin}, {Remigio}, {Armus},
  {Larson}, {D{\'\i}az-Santos}, {Evans}, {Stierwalt}, {Law}, {Malkan},
  {Linden}, {Song}, {van der Werf}, {Gao}, {Privon}, {Medling},
  {Barcos-Mu{\~n}oz}, {Hayward}, {Inami}, {Rich}, {Aalto}, {Appleton}, {Bohn},
  {B{\"o}ker}, {Brown}, {Charmandaris}, {Finnerty}, {Howell}, {Iwasawa},
  {Kemper}, {Marshall}, {Mazzarella}, {McKinney}, {Muller-Sanchez}, {Murphy},
  {Sanders}, \& {Surace}}]{GOALS_JWST_ngc7469}
{U}, V., {Lai}, T., {Bianchin}, M., {et~al.} 2022, \apjl, 940, L5

\bibitem[{{Weingartner} \& {Draine}(2001)}]{WeingartnerJ_01}
{Weingartner}, J.~C. \& {Draine}, B.~T. 2001, \apjs, 134, 263

\bibitem[{{Yamagishi} {et~al.}(2012){Yamagishi}, {Kaneda}, {Ishihara}, {Kondo},
  {Onaka}, {Suzuki}, \& {Minh}}]{YamagishiM_12}
{Yamagishi}, M., {Kaneda}, H., {Ishihara}, D., {et~al.} 2012, \aap, 541, A10

\bibitem[{Yang \& Li(2023)}]{yang2023aliphatics}
Yang, X. \& Li, A. 2023, The Astrophysical Journal Supplement Series, 268, 50

\bibitem[{{Yang} {et~al.}(2020){Yang}, {Li}, \& {Glaser}}]{Yang2020}
{Yang}, X.~J., {Li}, A., \& {Glaser}, R. 2020, \apjs, 247, 1

\bibitem[{{Yang} {et~al.}(2016{\natexlab{a}}){Yang}, {Li}, {Glaser}, \&
  {Zhong}}]{YangXJ_16}
{Yang}, X.~J., {Li}, A., {Glaser}, R., \& {Zhong}, J.~X. 2016{\natexlab{a}},
  \apj, 825, 22

\bibitem[{{Yang} {et~al.}(2016{\natexlab{b}}){Yang}, {Li}, {Glaser}, \&
  {Zhong}}]{Yang2016}
{Yang}, X.~J., {Li}, A., {Glaser}, R., \& {Zhong}, J.~X. 2016{\natexlab{b}},
  \apj, 825, 22

\bibitem[{{Yoast-Hull} {et~al.}(2013){Yoast-Hull}, {Everett}, {Gallagher}, \&
  {Zweibel}}]{Yoast_13_m82}
{Yoast-Hull}, T.~M., {Everett}, J.~E., {Gallagher}, J.~S., I., \& {Zweibel},
  E.~G. 2013, \apj, 768, 53

\bibitem[{{Yuan} {et~al.}(2019){Yuan}, {Murase}, \&
  {M{\'e}sz{\'a}ros}}]{Yuan_NGC3256}
{Yuan}, C., {Murase}, K., \& {M{\'e}sz{\'a}ros}, P. 2019, \apj, 878, 76

\end{thebibliography}
\bibliographystyle{aa}

\label{lastpage}
\end{document}